\documentclass[aps,pra,showpacs,twocolumn, longbibliography]{revtex4-1} %,longbibliography

\usepackage{graphicx}
\usepackage{dcolumn}
\usepackage{bm}
\usepackage{color}
\usepackage{amsmath}
\usepackage{amssymb}
\begin{document}

\title{Counter-propagating spectrally uncorrelated biphotons at 1550 nm generated from periodically poled $M$TiO$X$O$_4$ ($M$ = K, Rb, Cs; $X$ = P, As)}

\author{Wu-Hao Cai}
%\email{cwh827845249@gmail.com}
\author{Bei Wei}
%\email{wei.bei@foxmail.com}
%
\author{Shun Wang}
%%\email{shunwang@wit.edu.cn}
\author{Rui-Bo Jin}
\email{jrbqyj@gmail.com}
\affiliation{Hubei Key Laboratory of Optical Information and Pattern Recognition, Wuhan Institute of Technology, Wuhan 430205, PR China}

%% To be edited by editor
\date{\today }

%% To be edited by editor
% \doi{\url{https://doi.org/10.1364/JOSAB.401157}}

\begin{abstract}
We theoretically investigated spectrally uncorrelated biphotons generated in a counter-propagating spontaneous parametric downconversion (CP-SPDC) from periodically poled  $M$TiO$X$O$_4$ ($M$ = K, Rb, Cs; $X$ = P, As) crystals.
By numerical calculation, it was found that the five crystals from the KTP family can be used to generate heralded single photons with high spectral purity and wide tunability.
Under the type-0 phase-matching condition, the purity at 1550 nm was between 0.91 and 0.92, and the purity can be maintained over 0.90 from 1500 nm to 2000 nm wavelength.
Under the type-II phase-matching condition, the purity at 1550 nm was 0.96, 0.97, 0.97, 0.98, and 0.98 for PPKTP, PPRTP, PPKTA, PPRTA, and PPCTA, respectively; furthermore, the purity can be kept
over 0.96 for more than 600 nm wavelength range.
We also simulated the Hong--Ou--Mandel interference between independent photon sources for PPRTP crystals at 1550 nm, and interference visibility was 92\% (97\%) under type-0 (type-II) phase-matching condition.
This study may provide spectrally pure narrowband single-photon sources for quantum memories and quantum networks at telecom wavelengths.
\end{abstract}

\maketitle
\section{Introduction}
Biphotons generated from counter-propagating spontaneous parametric downconversion (CP-SPDC) offer a very unique photon source for quantum information processing (QIP). Compared with the conventional co-propagating SPDC (CO-SPDC) scheme, the CP-SPDC scheme has several merits.
First, the biphotons, i.e., the signal and the idler, can be easily separated, because the signal travels with the pump beam in the forward direction while the idler is propagating in the opposite direction. Therefore, the CP-SPDC scheme can take advantage of both maximal effective nonlinear coefficient and easy separation in type-0 phase-matching conditions \cite{Christ2009}.
Second, the bandwidth of the signal photons is very narrow, usually in the order of GHz, much lower than the typical THz level in CO-SPDC. Such a narrowband source is very useful for many
QIP applications that require GHz bandwidth to meet the need for quantum communication \cite{Li2016, Shi2020PRAppl} and quantum memory \cite{Bhaskar2020,  Zhong2017, Saglamyurek2015, Saglamyurek2016, Askarani2019}.
Conventionally, the methods of passive filtering \cite{Halder2007} and cavity-enhanced SPDC \cite{Scholz2009} are used to obtain photon with narrow bandwidth. However, these two methods usually reduce the brightness and stability, compared with the CP-SPDC scheme.
Third, by carefully designing the group-velocity matching (GVM) condition for CP-SPDC, the intrinsic spectral purity of the heralded single photons can be very high \cite{Christ2009, Gatti2018, Luo2020}. Such pure single photons can achieve high-visibility quantum interference without using any narrow bandpass filters \cite{Mosley2008PRL}.

Many theoretical and experimental works have been devoted to the study of CP-SPDC.
From the theoretical perspective:
In 1966, Harris proposed the first backward-wave oscillation scheme based on a three-wave-mixing processes \cite{Harris1966}.
In 2002, Booth et al. and Rossi et al. investigated the generation of biphoton from a CP-SPDC process in nonlinear media \cite{Booth2002, Rossi2002}.
In 2009, Christ et al. explored spectrally pure single-photon generation in a type-I phase-matched CP-SPDC in PPLN and PPKTP waveguides \cite{Christ2009}.
In 2011, Gong et al. proposed a scheme for the generation of counter-propagating polarization-entangled photons using a single dual-periodically poled KTP \cite{Gong2011}.
In 2015, Gatti et. al. analyzed the temporal coherence and correlation of counter-propagating twin photons in PPKTP and PPLN crystals \cite{Gatti2015}.
In the same year, Shukhin et al. simulated a fifth-order quasi-phase-matched CP-SPDC in a PPKTP
waveguide \cite{Shukhin2015}.
In 2017, Saravi et. al. proposed the generation of counter-propagating path-entangled photon pairs in a single periodic PPLN waveguide \cite{Saravi2017}.
In 2018, Gatti et al. studied the spectrally pure heralded-single-photon generation from CP-SPDC in a PPKTP crystal \cite{Gatti2018}.
From the experimental perspective: The first CP-SPDC experiment was realized in AlGaAs waveguide at 1550 nm by Lanco et al. in 2006 \cite{Lanco2006, Ravaro2005}.
However, the semiconductor materials have strong absorption at the telecom wavelengths, which limits the brightness of the biphoton source.
In 2007, Canalias et al. performed a counter-propagating optical parametric oscillation (OPO) experiment using a 1-mm-long PPKTP crystal with a poling period of 0.72 $\mu$m \cite{Canalias2007}. The wavelengths of the pump, signal and idler were 821, 1139 nm, and 2941 nm, respectively.
In 2019, Liu et al. experimentally demonstrated a 7.1 GHz (57 pm) narrowband entanglement sources at 1553 nm from a type-II phased matched third-order poled PPKTP crystal with a poling period of 1.3 $\mu$m \cite{Liu2019}.
In 2020, Luo et. al. reported on the generation of counter-propagating spectrally uncorrelated biphotons at 1550 nm in a 37-mm-long  type-0 phased-matched fifth-order-poled Ti-indiffused PPLN waveguide with a poling period of 1.7 $\mu$m \cite{Luo2020}.
By reviewing the previous important progress on CP-SPDC utilizing the material of AlGaAs, PPLN, or PPKTP \cite{Lanco2006, Ravaro2005, Canalias2007, Liu2019, Luo2020}, it can be noticed that the study on CP-SPDC is very promising, but it is still limited by the properties of the nonlinear material.
Therefore, it is still a field with high demand to develop novel nonlinear materials with higher nonlinear coefficient, higher transparency, higher spectral purity, and proper
poling period at telecom wavelengths.

The isomorphs of a KTP crystal, including  RTP ($\mathrm{RbTiOPO_4}$), KTA ($\mathrm{KTiOAsO_4}$), RTA ($\mathrm{RbTiOAsO_4}$), and CTA ($\mathrm{CsTiOAsO_4}$), have the general form of $M$TiO$X$O$_4$ with \{$M$ = K, Rb, Cs\} and \{$X$ = P, As (for $M$ = Cs only)\} \cite{Cheng1994, Shur2016}.
It was recently discovered that these four crystals have good performance in the type-II phase-matched CO-SPDC \cite{Jin2016PRAppl, Laudenbach2017, Jin2019OLT, Kim2015}.
These crystals still retain the desirable properties of their parent PPKTP crystal and can generate pure photons with high spectral purity (over 0.8) and wide tunability (from 1300 nm to 2100 nm) \cite{Jin2016PRAppl, Laudenbach2017}.
In this work, we explore the performance of the $M$TiO$X$O$_4$ in CP-SPDC under type-0 and type-II phase-matching conditions.

\section{Theory }
The biphoton state $\vert\psi\rangle$ generated from a CP-SPDC process can be written as
\begin{equation}\label{psi}
\vert\psi\rangle=\int_0^\infty\int_0^\infty\,\mathrm{d}\omega_s\,\mathrm{d}\omega_if(\omega_s,\omega_i)\hat a_s^\dag(\omega_s)\hat a_i^\dag(\omega_i)\vert0\rangle\vert0\rangle,
\end{equation}
where $\omega$ is the angular frequency, $\hat a^\dag$ is the creation operator, the subscripts \emph{s} and \emph{i} denote the signal and the idler photon, respectively. $f(\omega_s,\omega_i)$ is the joint spectral amplitude (JSA), which is the product of pump-envelope function (PEF) $\alpha(\omega_s,\omega_i)$ and phase-matching function (PMF) $\phi(\omega_s,\omega_i)$.
The PEF with a Gaussian-distribution can be written as \cite{Mosley2008NJP}
\begin{equation}\label{PEF}
\alpha(\omega_s, \omega_i)=\exp[-\frac{1}{2}\left(\frac{\omega_s+\omega_i-\omega_p}{\sigma_p}\right)^2],
\end{equation}
where $\sigma_p$ is the bandwidth of the pump, and the full-width at half-maximum (FWHM) in angular frequency is $FWHM_\omega=2\sqrt{ln(2)}\sigma_p \approx 1.67\sigma_p$.
The PEF can also be expressed with wavelength as the variable,
\begin{equation}\label{eq21}
\alpha(\lambda_s, \lambda_i)= \exp \left(  -\frac{1}{2} \left\{  \frac{{1/\lambda _s  + 1/\lambda _i  - 1/(\lambda _0 /2)}}{{\Delta \lambda /[(\lambda _0 /2)^2  - (\Delta \lambda /2)^2 ]}}  \right\}^2 \right),
\end{equation}
where  $\lambda _0 /2 $ is the central wavelength of the pump; the FWHM of the pump at intensity level is
FWHM$_\lambda$=$ \frac{2\sqrt{\ln (2)} {\lambda _0}^2   \Delta \lambda    \left({\lambda _0}^2-\Delta \lambda ^2\right)}{{\lambda _0}^4+\Delta \lambda ^4-2 {\lambda _0}^2 \Delta \lambda ^2 [1+\ln (4)]} $.
For $\Delta \lambda << \lambda _0$, FWHM$_\lambda\approx 2\sqrt{\ln(2)} \Delta \lambda  \approx 1.67\Delta \lambda $.

\begin{figure}[tbp]
\centering\includegraphics[width=6cm]{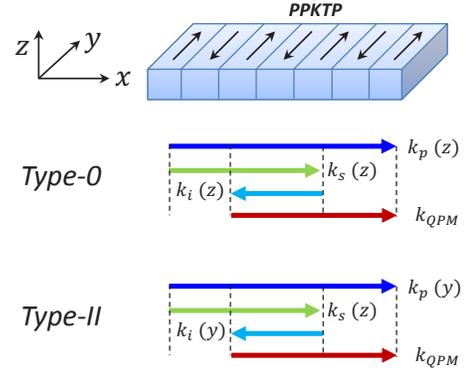}
\caption{Configurations of CP-SPDC under type-0 and type-II phase-matching conditions. The “x, y, z” axes indicate the direction of
polarizations.
 } \label{fig1}
\end{figure}

The PMF with a flat phase distribution can be described by
\begin{equation}\label{PMF}
\phi(\omega_s,\omega_i)=\operatorname{sinc}\left(\frac{\Delta kL}{2}\right),
\end{equation}
where \emph{L} is the length of the crystal and $\Delta k$ is the phase-mismatching vector.
In the CP-SPDC, the signal propagates in the same direction as the pump, while the idler propagates in the opposite direction, as shown in Fig. \ref{fig1}. $\Delta k$  can be written as
\begin{equation}\label{DeltaK}
\Delta k=k_s\color{red}-\color{black}k_i+k_{QPM}-k_p,
\end{equation}
 where $k_{QPM}={2 \pi}/{\Lambda}$ is the quasi-phase-matching (QPM) vector introduced by the periodically poled structure and $\Lambda$ is the poling period of the crystal. $k_{QPM}$ in a CP-SPDC scheme is generally much larger than the value in a CO-SPDC scheme, corresponding to a much shorter poling period $\Lambda$.

The  tilting  angle  $\theta$ between the ridge direction of the PMF and the positive direction of horizontal axis can be obtained as \cite{Jin2013}
\begin{equation}\label{theta}
\tan\theta=-\left( \frac{V_{g,p}^{-1}(\omega_p)-V_{g,s}^{-1}(\omega_s)}{V_{g,p}^{-1}(\omega_p)\color{red}+\color{black}V_{g,i}^{-1}(\omega_i)} \right),
\end{equation}
where $V_{g,\mu}=\frac{d\omega}{dk_\mu(\omega)}=\frac{1}{k_\mu^\prime(\omega)},(\mu=p, s, i)$ is the group velocity of the pump, the signal, and the idler photon.
When
\begin{equation}\label{gvm}
V_{g,p}^{-1}(\omega_p)=V_{g,s}^{-1}(\omega_s),
\end{equation}
the tilting angle $\theta$ equals to $0^{\circ}$, i.e. the corresponding PMF is distributed along the horizontal position. Under this condition, the purity of JSA can be maximized. Therefore, Eq. (\ref{gvm}) is called the GVM condition.
The spectral  purity can be calculated by applying Schmidt decomposition on the JSA \cite{Mosley2008NJP},
\begin{equation}\label{eqe1}
f(\omega _s ,\omega _i ) = \sum \limits_j  { c_j } \xi _j (\omega _s )\zeta _j (\omega _i ),
\end{equation}
where $\xi _j (\omega _s )$  and  $\zeta _j (\omega _i )$ are two orthogonal basis sets of spectral functions and $c_j$  is the normalized coefficient. The purity is defined as
\begin{equation}\label{eqe2}
p \equiv \sum \limits_j { c_j^4 } .
\end{equation}
The purity of a JSA is closely related with the tilting angle $\theta$.

In this work, we consider two types of phase-matching conditions for PPKTP crystal and its isomorphs: type-0 and type-II.
In the type-0 condition, all the signal, idler, and pump photons are polarized in the z directions, i.e.,
$\mathord{\buildrel{\lower3pt\hbox{$\scriptscriptstyle\rightharpoonup$}}
\over p} (z) \to \mathord{\buildrel{\lower3pt\hbox{$\scriptscriptstyle\rightharpoonup$}}
\over s} (z) + \mathord{\buildrel{\lower3pt\hbox{$\scriptscriptstyle\leftharpoonup$}}
\over i} (z)$,
 as shown in Fig. \ref{fig1}.
Under this configuration, the effective nonlinear coefficient can achieve the maximal value.
In the type-II condition, the pump and idler are polarized in the y direction, while the signal is polarized in the z direction,  i.e.,
$\mathord{\buildrel{\lower3pt\hbox{$\scriptscriptstyle\rightharpoonup$}}
\over p} (y) \to \mathord{\buildrel{\lower3pt\hbox{$\scriptscriptstyle\rightharpoonup$}}
\over s} (z) + \mathord{\buildrel{\lower3pt\hbox{$\scriptscriptstyle\leftharpoonup$}}
\over i} (y)$.
Under this configuration, the maximal purity can be achieved, as calculated in the next section.

\begin{figure}[tp]
\centering\includegraphics[width=8cm]{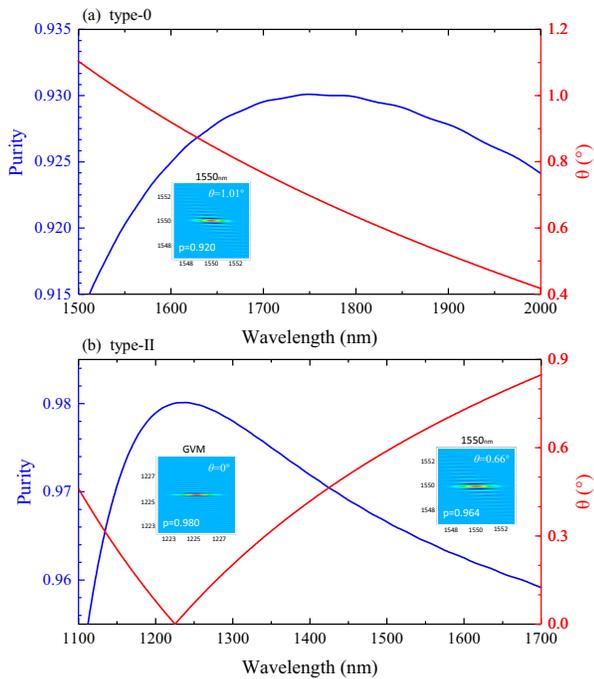}
\caption{Purity (blue line) and absolute value of tilt angle (red line) for the JSA of biphotons generated from PPKTP crystal under (a) type-0 and (b) type-II phase-matching conditions. The parameters of $L=5$ mm and $\Delta \lambda=0.16$ nm are
 adopted for (a);  $L=5$ mm  and $\Delta \lambda=0.20$ nm are used for (b). Insets show the JSA at (a)1550 nm, (b, left) 1225 nm, and (b, right) 1550nm.
 } \label{fig2}
\end{figure}

\begin{figure}[!htp]
\centering\includegraphics[width=7cm]{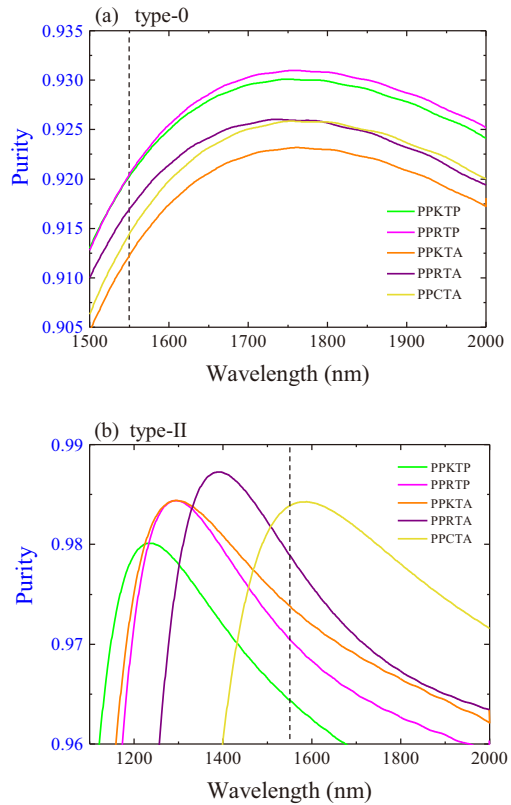}
\caption{
Purity for the JSA of biphotons generated from the isomorphs of PPKTP crystal under the (a) type-0 and (b) type-II phase-matching conditions.
The parameters of crystal pump bandwidth $\Delta\lambda$ and length $L$ ($\Delta\lambda$, $L$) for each crystal are as follows: for type-0 cases,
PPKTP (0.16 nm, 5 mm), PPRTP (0.16 nm, 5 mm), PPKTA (0.15 nm, 5 mm), PPKTA (0.18 nm, 4 mm) and PPCTA (0.18 nm, 4 mm );
for type-II cases,
PPKTP (0.2 nm, 5 mm), PPRTP (0.25 nm, 5 mm), PPKTA (0.25 nm, 5mm), PPKTA (0.25 nm, 7 mm) and PPCTA (0.2 nm, 10 mm).
 } \label{fig3}
\end{figure}

\begin{table*}[!htbp]
\centering\begin{tabular}{c|ccccc}
\hline \hline
Name                             &PPKTP                  &PPRTP                 &PPKTA                 &PPRTA                  &PPCTA               \\
Composition   &$\mathrm{KTiOPO_4}$    &$\mathrm{RbTiOPO_4}$  &$\mathrm{KTiOAsO_4}$  &$\mathrm{RbTiOAsO_4}$  &  $\mathrm{CsTiOAsO_4}$ \\
Sellmeier Eq.                  &\cite{Kato2002}        &\cite{Mikami2009}   &\cite{Feve2000}      &\cite{Kato2003}   & \cite{Mikami2011}
\\
 \hline
$\lambda_\textrm{GVM-0}$ (nm)     &2502.62                &2531.59               &2729.67               &2734.62                &2780.14                \\
$\Lambda_\textrm{GVM-0}$ (nm)      &686.266                &685.767               &734.277              &730.648                &728.149               \\
$\Lambda_\textrm{1550nm-0}$ (nm)   &419.637                &414.427               &410.937             &408.139                &399.928
\\
$p_\textrm{GVM-0}$                 &0.985                 &0.985                  &0.983                &0.983                 &0.983               \\
$p_\textrm{1550nm-0}$              &0.920                  &0.920                 &0.912                 &0.917                  &0.914                  \\
$d_\textrm{eff-0}$ (pm/V)          &9.5                    &9.7                   &9.6                   &9.8                    &11.2
\\
 \hline
$\lambda_\textrm{GVM-II}$ (nm)      &1225.19                &1282.04               &1284.84               &1379.66                &1577.17                \\
$\Lambda_\textrm{GVM-II}$ (nm)      &353.570                &363.835               &360.766               &383.846                &426.314               \\
$\Lambda_\textrm{1550nm-II}$ (nm)   &451.185                &442.863               &437.999               &432.963                &418.731
\\
$p_\textrm{GVM-II}$                 &0.980                  &0.984                 &0.984                 &0.987                  &0.984                  \\
$p_\textrm{1550nm-II}$              &0.964                  &0.970                 &0.974                 &0.979                  &0.984                  \\
$d_\textrm{eff-II}$ (pm/V)          &2.4                    &2.4                   &2.3                   &2.4                    &2.1                   \\
\hline \hline
\end{tabular}
\caption{\label{table1} Parameters of CP-SPDC under type-\textrm{0} and type-\textrm{II} phase-matching conditions for PPKTP  and  its isomorphs.
The reference of Sellmeier equations, the group-velocity-matched (GVM) wavelength ($\lambda_\textrm{GVM}$), the poling periods at GVM wavelength ($\Lambda_\textrm{GVM}$) and 1550 nm ($\Lambda_\textrm{1550 nm}$), purity at GVM wavelength ($p_\textrm{GVM}$) and 1550 nm ($p_\textrm{1550 nm}$), effective nonlinear coefficient ($d_\textrm{eff}$) are listed.
The  values  of $d_\textrm{eff-0}=d_{33}$ at 1550 nm and $d_\textrm{eff-II}=d_{32}$  at 1550 nm are taken from the SNLO v70 software package, developed by AS-Photonics, \emph{LLC} \cite{SNLO70}.
}
\end{table*}

\section{Calculation and Simulation}

First, we consider the calculation for PPKTP, and then we expand to the other four isomorphs. In all the calculation, we assume the wavelength is degenerated, i.e., $2\lambda_p = \lambda_s = \lambda_i$.
For PPKTP crystal under the type-0 phase-matching condition, the calculated GVM wavelength is at 2503 nm. This means the tilt angle $\theta$ is 0$^{\circ}$ at this wavelength. Then, we calculate $\theta$ for wavelengths from 1500 nm to 2000 nm. Surprisingly, the tilt angle only changes from 0.42$^{\circ}$ to 1.10$^{\circ}$, as shown by the red line in Fig. \ref{fig2}(a). Especially, the tilt angle is  1.01$^{\circ}$ at 1550 nm,  the most widely used wavelength for telecommunications.
Then, we optimize the crystal length $L$ and pump bandwidth $\Delta \lambda$ so as to maximize the spectral purity at 1550 nm.
The maximal purity is achieved at the value of 0.92, with the parameters of $L=5$ mm and $\Delta \lambda=0.16$ nm. %(FWHM$_\lambda$ = 0.27 nm)
The corresponding JSA is shown in the inset of Fig. \ref{fig2}(a). %, which is consistent with the results in Ref. (\cite{Christ2009}).
Further, by fixing the values of $L$ and  $\Delta \lambda$, we calculate the purity for other wavelengths. Calculation results show that the purity can maintain over 0.913 from 1500 nm to 2000 nm, as indicated by the blue line in Fig. \ref{fig2} (a).
Note that in the numerical calculation in this study, we use a grid size of 200*200 for all the JSAs.

Next, we consider the type-II phase-matching condition for PPKTP.
Under this condition, the calculated GVM wavelength is 1225 nm, corresponding to a tilting angle $\theta$ of 0$^{\circ}$.
Following a similar method, we also calculate the $\theta$ for wavelength from 1100 nm to 1700 nm. As indicated by the red line in Fig. \ref{fig2} (b), the tilt angle is changed from -0.46$^{\circ}$ to 0.85$^{\circ}$. Especially, $\theta$ is 0.66$^{\circ}$ at 1550 nm.
Then, we optimize the crystal length $L$ and pump bandwidth $\Delta \lambda$ to maximize the spectral purity for JSA at 1550 nm.
The calculated maximal purity is 0.96 at the parameters of $L=5$ mm and $\Delta \lambda=0.2$ nm. %(FWHM$_\lambda=0.33$ nm).
The corresponding JSA at 1550 nm is shown in the inset of Fig.\ref{fig2} (b).
Further, we calculate the purity for other wavelengths using the same parameters for $L$ and $\Delta \lambda$.
As indicated by the blue line in Fig. \ref{fig2} (b), the purity can be kept over 0.96 from 1100 nm to 1700 nm.
The JSA at the GVM wavelength 1225 nm is also shown in the inset of Fig. \ref{fig2} (b), with a purity as high as 0.98.

Following the same method, we calculate the case for the other four isomorph crystals.
Their GVM wavelengths, poling periods, and effective nonlinear coefficients are summarized in Table \ref{table1}.
Figure \ref{fig3} (a) shows the purity of the five crystals under a type-0 phase-matching condition. All the five crystals can keep the purity of over 0.90 from 1500 nm to 2000 nm. At 1550 nm wavelength, the purities are 0.92, 0.92, 0.91, 0.92, 0.92 for PPKTP, PPRTP, PPKTA, PPRTA, and PPCTA, respectively.
Figure \ref{fig3} (b) shows the purity of the five crystals under the  type-II phase-matching condition.
The purity is above 0.96 from 1100 nm to 1700 nm, and at 1550 nm, the purity is 0.96, 0.97, 0.97, 0.98, 0.98 for PPKTP, PPRTP, PPKTA, PPRTA, and PPCTA, respectively.
It can be concluded from Fig. \ref{fig3} that all four isomorphs possess similar properties as their parent crystal PPKTP.

\begin{figure*}[tp]
\centering\includegraphics[width=16cm]{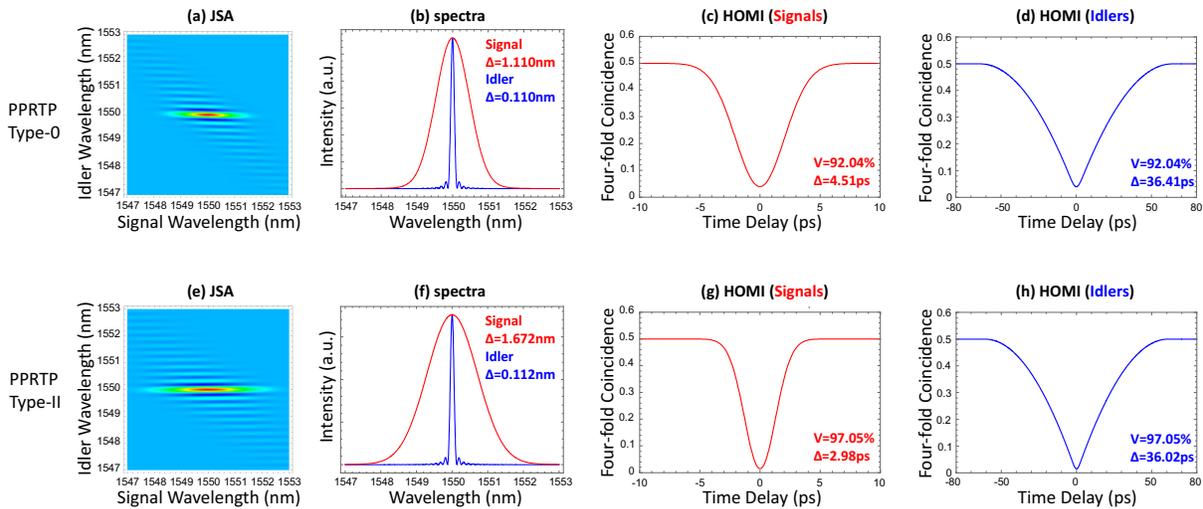}
\caption{(a, e) JSA, (b, f) spectra and (c), (d), (g), (h) HOM interference patterns for PPRTP under type-0 (up) and type-II (down) phase-matching conditions. The purity (p), FWHM of the spectra ($\Delta$), visibility (V), and FWHM of the HOM interference ($\Delta$) are shown in the figures.
The parameters of $L=5$ mm and $\Delta \lambda=0.16$ nm (FWHM = 0.27 nm) are
 adopted for the type-0 condition; $L=5$ mm  and $\Delta \lambda=0.25$ nm (FWHM = 0.42 nm) are used for the type-II condition.
 } \label{fig4}
\end{figure*}

The spectral purity of the single photons generated from CP-SPDC can be tested by a Hong--Ou--Mandel (HOM)  interference \cite{Hong1987} between two independent sources.
The typical experimental setup is shown in \cite{Mosley2008PRL, Jin2013PRA}, where two signals are sent to a 50:50 beam splitter and two idlers are detected to herald the arrival time of the signals.
Here we make a simulation of such a HOM interference for PPRTP  under type-0 and type-II phase-matching conditions.
The simulation is performed with the following equation \cite{Grice1997, Ou2007, Jin2020QUE}:
\begin{equation}\label{eq:P4}
\begin{split}
P_4 (\tau )  = & \frac{1}{4}  \int_0^\infty \int_0^\infty \int_0^\infty \int_0^\infty d\omega _{s_1} d\omega _{s_2} d\omega _{i_1} d\omega _{i_2}  \\ & {\rm{|}}f_1 (\omega _{s_1} ,\omega _{i_1} )f_2 (\omega _{s_2} ,\omega _{i_2} )- \\ & f_1 (\omega _{s_2} ,\omega _{i_1} )f_2 (\omega _{s_1} ,\omega _{i_2} )e^{ - i(\omega _{s_2}  - \omega _{s_1} )\tau } {\rm{|}}^{\rm{2}},
\end{split}
\end{equation}
where  $ P_4(\tau )$  is the fourfold coincidence probability, $\tau$ is the delay time,  the $f_{1(2)}$ is the JSA from the first (second) SPDC source.
Figure \ref{fig4} (a) shows the JSA of the biphotons generated from PPRTP under the type-0 phase-matching condition.
This JSA is obtained by using a pump laser with a pump bandwidth of  $\Delta \lambda=0.16$ nm   and a PPRTP with a length of $L = 5$ mm.
Figure \ref{fig4} (b) shows the spectral distribution of the signal and idler photons, obtained by projecting the joint spectral intensity onto the horizontal and vertical axes.
The signal (idler) has an FWHM of 1.11nm (0.11nm), i.e., the signal is about 10 times broader than the idler.
Figure \ref{fig4} (c) shows the HOM interference pattern of two signals heralded by two idlers with an FWHM of 4.51ps and a visibility of 92.04\%.
Figure \ref{fig4} (d) shows the case for two heralded idlers with an FWHM of 36.41ps  and a visibility of 92.04\%.
For comparison, we also simulate the case for PPRTP under type-II phase-matching condition, as shown in Figs. \ref{fig4}(e)--\ref{fig4}(h).
The JSA at 1550 nm in Fig. \ref{fig4} (e) has a purity of 0.97, which is obtained by using a pump bandwidth of  $\Delta \lambda=0.25$ nm  and a crystal length of $L = 5 $ mm.
The signal (idler) photon has an FWHM of 1.67 nm (0.11 nm) in Fig. \ref{fig4} (e),  i.e., the signal is about 15 times broader than the idler.
The HOM interference visibilities are 97.05\% in Figs. \ref{fig4} (g) and \ref{fig4} (h) , with the widths of 2.98 ps and 36.02 ps, respectively.
Comparing  Figs. \ref{fig4}(a)--\ref{fig4}(d) and Figs. \ref{fig4}(e)--\ref{fig4}(h),
it can be noticed that the spectral widths of the idler photons in the type-II phase-matching conditions are similar as the case in the type-0 condition. However, the purity and interference visibility are higher under the type-II  condition.

\section{Discussion}

Under the type-II phase-matching conditions, the CP-SPDC has two different subclasses:
$\mathord{\buildrel{\lower3pt\hbox{$\scriptscriptstyle\rightharpoonup$}}
\over p} (y) \to \mathord{\buildrel{\lower3pt\hbox{$\scriptscriptstyle\rightharpoonup$}}
\over s} (z) + \mathord{\buildrel{\lower3pt\hbox{$\scriptscriptstyle\leftharpoonup$}}
\over i} (y)$
and
$\mathord{\buildrel{\lower3pt\hbox{$\scriptscriptstyle\rightharpoonup$}}
\over p} (y) \to \mathord{\buildrel{\lower3pt\hbox{$\scriptscriptstyle\rightharpoonup$}}
\over s} (y) + \mathord{\buildrel{\lower3pt\hbox{$\scriptscriptstyle\leftharpoonup$}}
\over i} (z)$,
i.e, the signal may possess different polarizations \cite{Shukhin2015}.
The former case has been investigated in detail in this work.
In the latter case, however, the GVM wavelengths of $M$TiO$X$O$_4$ are far away from 1550 nm. As a result, the purity at 1550 nm is not as high as the value in the former case.
For example, in the latter case, the GVM wavelength of PPKTP is 2337 nm and the purity is 0.91, which is lower than the value of 0.964 in the former case (See Table \ref{table1}).
So, we did not focus on this subclass in this work.
Although the configuration in the latter case is not attractive for applications at the telecom wavelength, it is promising for applications at the middle-infrared wavelength.

The poling periods in Table \ref{table1} are all under 1 $\mu$m.
Currently, it is still technically challenging to fabricate poling period at the sub-micron scale.
Therefore, the simulation results in this work now can only be experimentally realized by using third- or fifth-order QPM method \cite{Liu2019, Luo2020}, which results in lower efficiency.
With the development of periodically poling technique, the technical difficulty may be overcome in the future.
Then, the spectrally uncorrelated counter-propagating biphotons in our simulation may be fully verified in the experiment.

In  Fig. \ref{fig4}, the purities at 1550 nm  are among 0.91 to  0.92 for the type-0 phase-matching condition,  and among 0.96 to 0.98 for the type-II phase-matching condition. The purities can be further improved to almost 1 by optimizing the arrangement of the poling period, for example by using  the recently developed custom poling technique based on quantum machine learning \cite{Cui2019PRAppl, Chen2019OE, Graffitti2018Optica}.

Cavity-enhanced SPDC with high performance crystals (such as PPLN) is also a good choice for narrowband and high-brightness sources, in spite of additional loss. If it is possible to adopt first-order QPM, the combination of CP-SPDC scheme and cavity enhancement should result in even higher spectral brightness due to less excited resonator modes.

For future applications, the narrowband single-photon source in our work may be stored in a quantum memory to realize large-scale quantum networks at telecom wavelengths.
More importantly,  the bandwidth of the source is tunable to satisfy the need of quantum memory without using any spectral filters, such as Fabry--Perot cavity or fibre Bragg grating.
For example, the bandwidth of the idler from a 5-mm-long PPRTP is 0.11 nm (13.74 GHz) in Fig. \ref{fig4} (b). By changing the crystal length from 1 mm to 30 mm, the bandwidth of the idler is tunable from 0.54 nm (67.43 GHz) to 0.019 nm (2.37 GHz). Such a tunable source is useful for the quantum memory of  erbium doped optical fibers \cite{Saglamyurek2015, Saglamyurek2016, Askarani2019}.

\section{Conclusion}
In conclusion, we have theoretically explored the performance of CP-SPDC in PPKTP, PPRTP, PPKTA, PPRTA, and PPCTA crystals at the telecom wavelengths.
Under the type-0 phase-matching condition of
 $\mathord{\buildrel{\lower3pt\hbox{$\scriptscriptstyle\rightharpoonup$}}
\over p} (z) \to \mathord{\buildrel{\lower3pt\hbox{$\scriptscriptstyle\rightharpoonup$}}
\over s} (z) + \mathord{\buildrel{\lower3pt\hbox{$\scriptscriptstyle\leftharpoonup$}}
\over i} (z)$,
the spectral purity is above 0.91 at 1550 nm for all crystals, and the purity can be maintained over 0.90 from 1500 nm to 2000 nm.
Under the type-II phase-matching condition of
$\mathord{\buildrel{\lower3pt\hbox{$\scriptscriptstyle\rightharpoonup$}}
\over p} (y) \to \mathord{\buildrel{\lower3pt\hbox{$\scriptscriptstyle\rightharpoonup$}}
\over s} (z) + \mathord{\buildrel{\lower3pt\hbox{$\scriptscriptstyle\leftharpoonup$}}
\over i} (y)$,
the purities of the five crystals are among 0.96 to 0.98 at 1550 nm, and the spectral purity is over 0.96 for more than 600 nm wavelength range.
To verify the spectral purity, we take PPRTP as an example to simulate the HOM interference between independent photon sources.
The heralded signals (idlers) shows a bandwidth of 1.11 nm (0.11nm) under the type-0 phase-matching condition, and a bandwidth of 1.67 nm (0.11 nm) under the type-II phase-matching condition. The interference visibility of 92.04\% (97.05\%) can be achieved for the type-0 (type-II) phase-matching condition.
This study may provide narrowband single photons with high spectral purity for quantum memory and quantum networks at telecom wavelengths.

\section*{Acknowledgments}
This work is partially supported by the National Natural Science Foundations of China (Grant Nos.91836102, 11704290). We thank Prof. Zheshen Zhang, Prof. Yan-Xiao Gong and Dr. Yi-Chen Liu for helpful discussions.


\begin{thebibliography}{43}%
\makeatletter
\providecommand \@ifxundefined [1]{%
 \@ifx{#1\undefined}
}%
\providecommand \@ifnum [1]{%
 \ifnum #1\expandafter \@firstoftwo
 \else \expandafter \@secondoftwo
 \fi
}%
\providecommand \@ifx [1]{%
 \ifx #1\expandafter \@firstoftwo
 \else \expandafter \@secondoftwo
 \fi
}%
\providecommand \natexlab [1]{#1}%
\providecommand \enquote  [1]{``#1''}%
\providecommand \bibnamefont  [1]{#1}%
\providecommand \bibfnamefont [1]{#1}%
\providecommand \citenamefont [1]{#1}%
\providecommand \href@noop [0]{\@secondoftwo}%
\providecommand \href [0]{\begingroup \@sanitize@url \@href}%
\providecommand \@href[1]{\@@startlink{#1}\@@href}%
\providecommand \@@href[1]{\endgroup#1\@@endlink}%
\providecommand \@sanitize@url [0]{\catcode `\\12\catcode `\$12\catcode
  `\&12\catcode `\#12\catcode `\^12\catcode `\_12\catcode `\%12\relax}%
\providecommand \@@startlink[1]{}%
\providecommand \@@endlink[0]{}%
\providecommand \url  [0]{\begingroup\@sanitize@url \@url }%
\providecommand \@url [1]{\endgroup\@href {#1}{\urlprefix }}%
\providecommand \urlprefix  [0]{URL }%
\providecommand \Eprint [0]{\href }%
\providecommand \doibase [0]{http://dx.doi.org/}%
\providecommand \selectlanguage [0]{\@gobble}%
\providecommand \bibinfo  [0]{\@secondoftwo}%
\providecommand \bibfield  [0]{\@secondoftwo}%
\providecommand \translation [1]{[#1]}%
\providecommand \BibitemOpen [0]{}%
\providecommand \bibitemStop [0]{}%
\providecommand \bibitemNoStop [0]{.\EOS\space}%
\providecommand \EOS [0]{\spacefactor3000\relax}%
\providecommand \BibitemShut  [1]{\csname bibitem#1\endcsname}%
\let\auto@bib@innerbib\@empty
%</preamble>
\bibitem [{\citenamefont {Christ}\ \emph {et~al.}(2009)\citenamefont {Christ},
  \citenamefont {Eckstein}, \citenamefont {Mosley},\ and\ \citenamefont
  {Silberhorn}}]{Christ2009}%
  \BibitemOpen
  \bibfield  {author} {\bibinfo {author} {\bibfnamefont {A.}~\bibnamefont
  {Christ}}, \bibinfo {author} {\bibfnamefont {A.}~\bibnamefont {Eckstein}},
  \bibinfo {author} {\bibfnamefont {P.~J.}\ \bibnamefont {Mosley}}, \ and\
  \bibinfo {author} {\bibfnamefont {C.}~\bibnamefont {Silberhorn}},\ }\bibfield
   {title} {\enquote {\bibinfo {title} {Pure single photon generation by
  type-{I} {PDC} with backward-wave amplification},}\ }\href {\doibase
  10.1364/oe.17.003441} {\bibfield  {journal} {\bibinfo  {journal} {Opt.
  Express}\ }\textbf {\bibinfo {volume} {17}},\ \bibinfo {pages} {3441-3446}
  (\bibinfo {year} {2009})}\BibitemShut {NoStop}%
\bibitem [{\citenamefont {Li}\ \emph {et~al.}(2016)\citenamefont {Li},
  \citenamefont {Zhou}, \citenamefont {Xu}, \citenamefont {Xu}, \citenamefont
  {Shi},\ and\ \citenamefont {Guo}}]{Li2016}%
  \BibitemOpen
  \bibfield  {author} {\bibinfo {author} {\bibfnamefont {Y.-H.}\ \bibnamefont
  {Li}}, \bibinfo {author} {\bibfnamefont {Z.-Y.}\ \bibnamefont {Zhou}},
  \bibinfo {author} {\bibfnamefont {Z.-H.}\ \bibnamefont {Xu}}, \bibinfo
  {author} {\bibfnamefont {L.-X.}\ \bibnamefont {Xu}}, \bibinfo {author}
  {\bibfnamefont {B.-S.}\ \bibnamefont {Shi}}, \ and\ \bibinfo {author}
  {\bibfnamefont {G.-C.}\ \bibnamefont {Guo}},\ }\bibfield  {title}
  {\enquote {\bibinfo {title} {Multiplexed entangled photon-pair sources for
  all-fiber quantum networks},}\ }\href@noop {} {\bibfield  {journal} {\bibinfo
   {journal} {Phys. Rev. A}\ }\textbf {\bibinfo {volume} {94}},\ \bibinfo {pages} {043810} (\bibinfo {year}
  {2016})}\BibitemShut {NoStop}%
\bibitem [{\citenamefont {Shi}\ \emph {et~al.}(2020)\citenamefont {Shi},
  \citenamefont {Zhang},\ and\ \citenamefont {Zhuang}}]{Shi2020PRAppl}%
  \BibitemOpen
  \bibfield  {author} {\bibinfo {author} {\bibfnamefont {H.}\ \bibnamefont
  {Shi}}, \bibinfo {author} {\bibfnamefont {Z.}\ \bibnamefont {Zhang}}, \
  and\ \bibinfo {author} {\bibfnamefont {Q.}\ \bibnamefont {Zhuang}},\
  }\bibfield  {title} {\enquote {\bibinfo {title} {Practical route to
  entanglement-assisted communication over noisy bosonic channels},}\
  }\href@noop {} {\bibfield  {journal} {\bibinfo  {journal} {Phys. Rev.
  Applied}\ }\textbf {\bibinfo {volume} {13}},\ \bibinfo {pages} {034029}
  (\bibinfo {year} {2020})}\BibitemShut {NoStop}%
\bibitem [{\citenamefont {Bhaskar}\ \emph {et~al.}(2020)\citenamefont
  {Bhaskar}, \citenamefont {Riedinger}, \citenamefont {Machielse},
  \citenamefont {Levonian}, \citenamefont {Nguyen}, \citenamefont {Knall},
  \citenamefont {Park}, \citenamefont {Englund}, \citenamefont {Lon?ar},
  \citenamefont {Sukachev},\ and\ \citenamefont {Lukin}}]{Bhaskar2020}%
  \BibitemOpen
  \bibfield  {author} {\bibinfo {author} {\bibfnamefont {M.~K.}\ \bibnamefont
  {Bhaskar}}, \bibinfo {author} {\bibfnamefont {R.}~\bibnamefont {Riedinger}},
  \bibinfo {author} {\bibfnamefont {B.}~\bibnamefont {Machielse}}, \bibinfo
  {author} {\bibfnamefont {D.~S.}\ \bibnamefont {Levonian}}, \bibinfo {author}
  {\bibfnamefont {C.~T.}\ \bibnamefont {Nguyen}}, \bibinfo {author}
  {\bibfnamefont {E.~N.}\ \bibnamefont {Knall}}, \bibinfo {author}
  {\bibfnamefont {H.}~\bibnamefont {Park}}, \bibinfo {author} {\bibfnamefont
  {D.}~\bibnamefont {Englund}}, \bibinfo {author} {\bibfnamefont
  {M.}~\bibnamefont {Loncar}}, \bibinfo {author} {\bibfnamefont {D.~D.}\
  \bibnamefont {Sukachev}}, \ and\ \bibinfo {author} {\bibfnamefont {M.~D.}\
  \bibnamefont {Lukin}},\ }\bibfield  {title} {\enquote {\bibinfo {title}
  {Experimental demonstration of memory-enhanced quantum communication},}\
  }\href@noop {} {\bibfield  {journal} {\bibinfo  {journal} {Nature}\ }\textbf
  {\bibinfo {volume} {580}},\ \bibinfo {pages} {60--64} (\bibinfo {year}
  {2020})}\BibitemShut {NoStop}%
\bibitem [{\citenamefont {Zhong}\ \emph {et~al.}(2017)\citenamefont {Zhong},
  \citenamefont {Kindem}, \citenamefont {Rochman},\ and\ \citenamefont
  {Faraon}}]{Zhong2017}%
  \BibitemOpen
  \bibfield  {author} {\bibinfo {author} {\bibfnamefont {T.}\ \bibnamefont
  {Zhong}}, \bibinfo {author} {\bibfnamefont {J. M.}\ \bibnamefont
  {Kindem}}, \bibinfo {author} {\bibfnamefont {J.}\ \bibnamefont {Rochman}},
  \ and\ \bibinfo {author} {\bibfnamefont {A.}\ \bibnamefont {Faraon}},\
  }\bibfield  {title} {\enquote {\bibinfo {title} {Interfacing broadband
  photonic qubits to on-chip cavity-protected rare-earth ensembles},}\
  }\href@noop {} {\bibfield  {journal} {\bibinfo  {journal} {Nat. Commun.}\
  }\textbf {\bibinfo {volume} {8}},\ \bibinfo {pages} {14107} (\bibinfo {year} {2017})}\BibitemShut
  {NoStop}%
\bibitem [{\citenamefont {Saglamyurek}\ \emph {et~al.}(2015)\citenamefont
  {Saglamyurek}, \citenamefont {Jin}, \citenamefont {Verma}, \citenamefont
  {Shaw}, \citenamefont {Marsili}, \citenamefont {Nam}, \citenamefont {Oblak},\
  and\ \citenamefont {Tittel}}]{Saglamyurek2015}%
  \BibitemOpen
  \bibfield  {author} {\bibinfo {author} {\bibfnamefont {E.}\ \bibnamefont
  {Saglamyurek}}, \bibinfo {author} {\bibfnamefont {J.}\ \bibnamefont
  {Jin}}, \bibinfo {author} {\bibfnamefont {V. B.}\ \bibnamefont {Verma}},
  \bibinfo {author} {\bibfnamefont {M. D.}\ \bibnamefont {Shaw}}, \bibinfo
  {author} {\bibfnamefont {F.}\ \bibnamefont {Marsili}}, \bibinfo
  {author} {\bibfnamefont {S. W.}\ \bibnamefont {Nam}}, \bibinfo {author}
  {\bibfnamefont {D.}\ \bibnamefont {Oblak}}, \ and\ \bibinfo {author}
  {\bibfnamefont {W.}\ \bibnamefont {Tittel}},\ }\bibfield  {title}
  {\enquote {\bibinfo {title} {Quantum storage of entangled telecom-wavelength
  photons in an erbium-doped optical fibre},}\ }\href@noop {} {\bibfield
  {journal} {\bibinfo  {journal} {Nat. Photonics}\ }\textbf {\bibinfo {volume}
  {9}},\ \bibinfo {pages} {83--87} (\bibinfo {year} {2015})}\BibitemShut
  {NoStop}%
\bibitem [{\citenamefont {Saglamyurek}\ \emph {et~al.}(2016)\citenamefont
  {Saglamyurek}, \citenamefont {Puigibert}, \citenamefont {Zhou}, \citenamefont
  {Giner}, \citenamefont {Marsili}, \citenamefont {Verma}, \citenamefont {Nam},
  \citenamefont {Oesterling}, \citenamefont {Nippa}, \citenamefont {Oblak},\
  and\ \citenamefont {Tittel}}]{Saglamyurek2016}%
  \BibitemOpen
  \bibfield  {author} {\bibinfo {author} {\bibfnamefont {E.}\ \bibnamefont
  {Saglamyurek}}, \bibinfo {author} {\bibfnamefont {M. G.}\
  \bibnamefont {Puigibert}}, \bibinfo {author} {\bibfnamefont {Q.}\
  \bibnamefont {Zhou}}, \bibinfo {author} {\bibfnamefont {L.}\
  \bibnamefont {Giner}}, \bibinfo {author} {\bibfnamefont {F.}\
  \bibnamefont {Marsili}}, \bibinfo {author} {\bibfnamefont {V. B.}\
  \bibnamefont {Verma}}, \bibinfo {author} {\bibfnamefont {S. W.}\
  \bibnamefont {Nam}}, \bibinfo {author} {\bibfnamefont {L.}\ \bibnamefont
  {Oesterling}}, \bibinfo {author} {\bibfnamefont {D.}\ \bibnamefont
  {Nippa}}, \bibinfo {author} {\bibfnamefont {D.}\ \bibnamefont {Oblak}}, \
  and\ \bibinfo {author} {\bibfnamefont {W.}\ \bibnamefont {Tittel}},\
  }\bibfield  {title} {\enquote {\bibinfo {title} {A multiplexed light-matter
  interface for fibre-based quantum networks},}\ }\href@noop {} {\bibfield
  {journal} {\bibinfo  {journal} {Nat. Commun.}\ }\textbf {\bibinfo {volume}
  {7}},\ \bibinfo {pages} {11202} (\bibinfo {year} {2016})}\BibitemShut {NoStop}%
\bibitem [{\citenamefont {Askarani}\ \emph {et~al.}(2019)\citenamefont
  {Askarani}, \citenamefont {Puigibert}, \citenamefont {Lutz}, \citenamefont
  {Verma}, \citenamefont {Shaw}, \citenamefont {Nam}, \citenamefont {Sinclair},
  \citenamefont {Oblak},\ and\ \citenamefont {Tittel}}]{Askarani2019}%
  \BibitemOpen
  \bibfield  {author} {\bibinfo {author} {\bibfnamefont {M. F.}\
  \bibnamefont {Askarani}}, \bibinfo {author} {\bibfnamefont
  {M. G.}\ \bibnamefont {Puigibert}}, \bibinfo {author}
  {\bibfnamefont {T.}\ \bibnamefont {Lutz}}, \bibinfo {author}
  {\bibfnamefont {V. B.}\ \bibnamefont {Verma}}, \bibinfo {author}
  {\bibfnamefont {M. D.}\ \bibnamefont {Shaw}}, \bibinfo {author}
  {\bibfnamefont {S. W.}\ \bibnamefont {Nam}}, \bibinfo {author}
  {\bibfnamefont {N.}\ \bibnamefont {Sinclair}}, \bibinfo {author}
  {\bibfnamefont {D.}\ \bibnamefont {Oblak}}, \ and\ \bibinfo {author}
  {\bibfnamefont {W.}\ \bibnamefont {Tittel}},\ }\bibfield  {title}
  {\enquote {\bibinfo {title} {Storage and reemission of heralded
  telecommunication-wavelength photons using a crystal waveguide},}\
  }\href@noop {} {\bibfield  {journal} {\bibinfo  {journal} {Phys. Rev. Appl.}\
  }\textbf {\bibinfo {volume} {11}},\ \bibinfo {pages} {054056} (\bibinfo {year} {2019})}\BibitemShut
  {NoStop}%
\bibitem [{\citenamefont {Halder}\ \emph {et~al.}(2007)\citenamefont {Halder},
  \citenamefont {Beveratos}, \citenamefont {Gisin}, \citenamefont {Scarani},
  \citenamefont {Simon},\ and\ \citenamefont {Zbinden}}]{Halder2007}%
  \BibitemOpen
  \bibfield  {author} {\bibinfo {author} {\bibfnamefont {M.}\
  \bibnamefont {Halder}}, \bibinfo {author} {\bibfnamefont {A.}\
  \bibnamefont {Beveratos}}, \bibinfo {author} {\bibfnamefont {N.}\
  \bibnamefont {Gisin}}, \bibinfo {author} {\bibfnamefont {V.}\
  \bibnamefont {Scarani}}, \bibinfo {author} {\bibfnamefont {C.}\
  \bibnamefont {Simon}}, \ and\ \bibinfo {author} {\bibfnamefont {H.}\
  \bibnamefont {Zbinden}},\ }\bibfield  {title} {\enquote {\bibinfo {title}
  {Entangling independent photons by time~measurement},}\ }\href {\doibase
  10.1038/nphys700} {\bibfield  {journal} {\bibinfo  {journal} {Nat. Phys.}\
  }\textbf {\bibinfo {volume} {3}},\ \bibinfo {pages} {692--695} (\bibinfo
  {year} {2007})}\BibitemShut {NoStop}%
\bibitem [{\citenamefont {Scholz}\ \emph {et~al.}(2009)\citenamefont {Scholz},
  \citenamefont {Koch},\ and\ \citenamefont {Benson}}]{Scholz2009}%
  \BibitemOpen
  \bibfield  {author} {\bibinfo {author} {\bibfnamefont {M.}\
  \bibnamefont {Scholz}}, \bibinfo {author} {\bibfnamefont {L.}\ \bibnamefont
  {Koch}}, \ and\ \bibinfo {author} {\bibfnamefont {O.}\ \bibnamefont
  {Benson}},\ }\bibfield  {title} {\enquote {\bibinfo {title} {Statistics of
  narrow-band single photons for quantum memories generated by ultrabright
  cavity-enhanced parametric down-conversion},}\ }\href@noop {} {\bibfield
  {journal} {\bibinfo  {journal} {Phys. Rev. Lett.}\ }\textbf {\bibinfo
  {volume} {102}},\ \bibinfo {pages} {063603} (\bibinfo {year} {2009})}\BibitemShut {NoStop}%
\bibitem [{\citenamefont {Gatti}\ and\ \citenamefont
  {Brambilla}(2018)}]{Gatti2018}%
  \BibitemOpen
  \bibfield  {author} {\bibinfo {author} {\bibfnamefont {A.}\
  \bibnamefont {Gatti}}\ and\ \bibinfo {author} {\bibfnamefont {E.}\
  \bibnamefont {Brambilla}},\ }\bibfield  {title} {\enquote {\bibinfo {title}
  {Heralding pure single photons: A comparison between counterpropagating and
  copropagating twin photons},}\ }\href@noop {} {\bibfield  {journal} {\bibinfo
   {journal} {Phys. Rev. A}\ }\textbf {\bibinfo {volume} {97}},\ \bibinfo {pages} {013838} (\bibinfo {year}
  {2018})}\BibitemShut {NoStop}%
\bibitem [{\citenamefont {Luo}\ \emph {et~al.}(2020)\citenamefont {Luo},
  \citenamefont {Ansari}, \citenamefont {Massaro}, \citenamefont {Santandrea},
  \citenamefont {Eigner}, \citenamefont {Ricken}, \citenamefont {Herrmann},\
  and\ \citenamefont {Silberhorn}}]{Luo2020}%
  \BibitemOpen
  \bibfield  {author} {\bibinfo {author} {\bibfnamefont {K.-H.}\
  \bibnamefont {Luo}}, \bibinfo {author} {\bibfnamefont {V.}\ \bibnamefont
  {Ansari}}, \bibinfo {author} {\bibfnamefont {M.}\ \bibnamefont
  {Massaro}}, \bibinfo {author} {\bibfnamefont {M.}\ \bibnamefont
  {Santandrea}}, \bibinfo {author} {\bibfnamefont {C.}\ \bibnamefont
  {Eigner}}, \bibinfo {author} {\bibfnamefont {R.}\ \bibnamefont
  {Ricken}}, \bibinfo {author} {\bibfnamefont {H.}\ \bibnamefont
  {Herrmann}}, \ and\ \bibinfo {author} {\bibfnamefont {C.}\
  \bibnamefont {Silberhorn}},\ }\bibfield  {title} {\enquote {\bibinfo {title}
  {Counter-propagating photon pair generation in a nonlinear waveguide},}\
  }\href {http://www.opticsexpress.org/abstract.cfm?URI=oe-28-3-3215}
  {\bibfield  {journal} {\bibinfo  {journal} {Opt. Express}\ }\textbf {\bibinfo
  {volume} {28}},\ \bibinfo {pages} {3215--3225} (\bibinfo {year}
  {2020})}\BibitemShut {NoStop}%
\bibitem [{\citenamefont {Mosley}\ \emph
  {et~al.}(2008{\natexlab{a}})\citenamefont {Mosley}, \citenamefont {Lundeen},
  \citenamefont {Smith}, \citenamefont {Wasylczyk}, \citenamefont {U'Ren},
  \citenamefont {Silberhorn},\ and\ \citenamefont {Walmsley}}]{Mosley2008PRL}%
  \BibitemOpen
  \bibfield  {author} {\bibinfo {author} {\bibfnamefont {P. J.}\
  \bibnamefont {Mosley}}, \bibinfo {author} {\bibfnamefont {J. S.}\
  \bibnamefont {Lundeen}}, \bibinfo {author} {\bibfnamefont {B. J.}\
  \bibnamefont {Smith}}, \bibinfo {author} {\bibfnamefont {P.}\ \bibnamefont
  {Wasylczyk}}, \bibinfo {author} {\bibfnamefont {A. B.}\ \bibnamefont
  {U'Ren}}, \bibinfo {author} {\bibfnamefont {C.}\ \bibnamefont
  {Silberhorn}}, \ and\ \bibinfo {author} {\bibfnamefont {I. A.}\ \bibnamefont
  {Walmsley}},\ }\bibfield  {title} {\enquote {\bibinfo {title} {Heralded
  generation of ultrafast single photons in pure quantum states},}\ }\href@noop
  {} {\bibfield  {journal} {\bibinfo  {journal} {Phys. Rev. Lett.}\ }\textbf
  {\bibinfo {volume} {100}},\ \bibinfo {pages} {133601} (\bibinfo {year}
  {2008}{\natexlab{a}})}\BibitemShut {NoStop}%
\bibitem [{\citenamefont {Harris}(1966)}]{Harris1966}%
  \BibitemOpen
  \bibfield  {author} {\bibinfo {author} {\bibfnamefont {S.~E.}\ \bibnamefont
  {Harris}},\ }\bibfield  {title} {\enquote {\bibinfo {title} {Proposed
  backward wave oscillation in the infrared},}\ }\href {\doibase
  10.1063/1.1754668} {\bibfield  {journal} {\bibinfo  {journal} {Appl. Phys.
  Lett.}\ }\textbf {\bibinfo {volume} {9}},\ \bibinfo {pages} {114--116}
  (\bibinfo {year} {1966})}\BibitemShut {NoStop}%
\bibitem [{\citenamefont {Booth}\ \emph {et~al.}(2002)\citenamefont {Booth},
  \citenamefont {Atature}, \citenamefont {Giuseppe}, \citenamefont {Saleh},
  \citenamefont {Sergienko},\ and\ \citenamefont {Teich}}]{Booth2002}%
  \BibitemOpen
  \bibfield  {author} {\bibinfo {author} {\bibfnamefont {M. C.}\ \bibnamefont
  {Booth}}, \bibinfo {author} {\bibfnamefont {M.}\ \bibnamefont {Atature}},
  \bibinfo {author} {\bibfnamefont {G. D.}\ \bibnamefont {Giuseppe}},
  \bibinfo {author} {\bibfnamefont {B. E. A.}\ \bibnamefont {Saleh}},
  \bibinfo {author} {\bibfnamefont {A. V.}\ \bibnamefont {Sergienko}}, \
  and\ \bibinfo {author} {\bibfnamefont {M. C.}\ \bibnamefont {Teich}},\
  }\bibfield  {title} {\enquote {\bibinfo {title} {Counterpropagating entangled
  photons from a waveguide with periodic nonlinearity},}\ }\href@noop {}
  {\bibfield  {journal} {\bibinfo  {journal} {Phys. Rev. A}\ }\textbf {\bibinfo
  {volume} {66}},\ \bibinfo {pages} {023815} (\bibinfo {year} {2002})}\BibitemShut {NoStop}%
\bibitem [{\citenamefont {Rossi}\ and\ \citenamefont
  {Berger}(2002)}]{Rossi2002}%
  \BibitemOpen
  \bibfield  {author} {\bibinfo {author} {\bibfnamefont {A. D.}\ \bibnamefont
  {Rossi}}\ and\ \bibinfo {author} {\bibfnamefont {V.}~\bibnamefont {Berger}},\
  }\bibfield  {title} {\enquote {\bibinfo {title} {Counterpropagating twin
  photons by parametric fluorescence},}\ }\href@noop {} {\bibfield  {journal}
  {\bibinfo  {journal} {Phys. Rev. Lett.}\ }\textbf {\bibinfo {volume} {88}},\ \bibinfo {pages} {043901}
  (\bibinfo {year} {2002})}\BibitemShut {NoStop}%
\bibitem [{\citenamefont {Gong}\ \emph {et~al.}(2011)\citenamefont {Gong},
  \citenamefont {Xie}, \citenamefont {Xu}, \citenamefont {Yu}, \citenamefont
  {Xue},\ and\ \citenamefont {Zhu}}]{Gong2011}%
  \BibitemOpen
  \bibfield  {author} {\bibinfo {author} {\bibfnamefont {Y.-X.}\
  \bibnamefont {Gong}}, \bibinfo {author} {\bibfnamefont {Z.-D.}\
  \bibnamefont {Xie}}, \bibinfo {author} {\bibfnamefont {P.}\ \bibnamefont
  {Xu}}, \bibinfo {author} {\bibfnamefont {X.-Q.}\ \bibnamefont {Yu}},
  \bibinfo {author} {\bibfnamefont {P.}\ \bibnamefont {Xue}}, \ and\ \bibinfo
  {author} {\bibfnamefont {S.-N.}\ \bibnamefont {Zhu}},\ }\bibfield  {title}
  {\enquote {\bibinfo {title} {Compact source of narrow-band counterpropagating
  polarization-entangled photon pairs using a single dual-periodically-poled
  crystal},}\ }\href@noop {} {\bibfield  {journal} {\bibinfo  {journal} {Phys.
  Rev. A}\ }\textbf {\bibinfo {volume} {84}},\ \bibinfo {053825} {054056} (\bibinfo {year}
  {2011})}\BibitemShut {NoStop}%
\bibitem [{\citenamefont {Gatti}\ \emph {et~al.}(2015)\citenamefont {Gatti},
  \citenamefont {Corti},\ and\ \citenamefont {Brambilla}}]{Gatti2015}%
  \BibitemOpen
  \bibfield  {author} {\bibinfo {author} {\bibfnamefont {A.}~\bibnamefont
  {Gatti}}, \bibinfo {author} {\bibfnamefont {T.}~\bibnamefont {Corti}}, \ and\
  \bibinfo {author} {\bibfnamefont {E.}~\bibnamefont {Brambilla}},\ }\bibfield
  {title} {\enquote {\bibinfo {title} {Temporal coherence and correlation of
  counterpropagating twin photons},}\ }\href@noop {} {\bibfield  {journal}
  {\bibinfo  {journal} {Phys. Rev. A}\ }\textbf {\bibinfo {volume} {92}},\
  \bibinfo {pages} {053809} (\bibinfo {year} {2015})}\BibitemShut {NoStop}%
\bibitem [{\citenamefont {Shukhin}\ \emph {et~al.}(2015)\citenamefont
  {Shukhin}, \citenamefont {Akatiev}, \citenamefont {Latypov}, \citenamefont
  {Shkalikov},\ and\ \citenamefont {Kalachev}}]{Shukhin2015}%
  \BibitemOpen
  \bibfield  {author} {\bibinfo {author} {\bibfnamefont {A. A}\ \bibnamefont
  {Shukhin}}, \bibinfo {author} {\bibfnamefont {D. O}\ \bibnamefont {Akatiev}},
  \bibinfo {author} {\bibfnamefont {I. Z}\ \bibnamefont {Latypov}}, \bibinfo
  {author} {\bibfnamefont {A. V}\ \bibnamefont {Shkalikov}}, \ and\ \bibinfo
  {author} {\bibfnamefont {A. A}\ \bibnamefont {Kalachev}},\ }\bibfield  {title}
  {\enquote {\bibinfo {title} {Simulating single-photon sources based on
  backward-wave spontaneous parametric down-conversion in a periodically poled
  {KTP} waveguide},}\ }\href {\doibase 10.1088/1742-6596/613/1/012015}
  {\bibfield  {journal} {\bibinfo  {journal} {J. Phys.: Conf. Ser.}\ }\textbf
  {\bibinfo {volume} {613}},\ \bibinfo {pages} {012015} (\bibinfo {year}
  {2015})}\BibitemShut {NoStop}%
\bibitem [{\citenamefont {Saravi}\ \emph {et~al.}(2017)\citenamefont {Saravi},
  \citenamefont {Pertsch},\ and\ \citenamefont {Setzpfandt}}]{Saravi2017}%
  \BibitemOpen
  \bibfield  {author} {\bibinfo {author} {\bibfnamefont {S.}\ \bibnamefont
  {Saravi}}, \bibinfo {author} {\bibfnamefont {T.}\ \bibnamefont
  {Pertsch}}, \ and\ \bibinfo {author} {\bibfnamefont {F.}\ \bibnamefont
  {Setzpfandt}},\ }\bibfield  {title} {\enquote {\bibinfo {title} {Generation
  of counterpropagating path-entangled photon pairs in a single periodic
  waveguide},}\ }\href@noop {} {\bibfield  {journal} {\bibinfo  {journal}
  {Phys. Rev. Lett.}\ }\textbf {\bibinfo {volume} {118}},\ \bibinfo {pages} {183603} (\bibinfo {year}
  {2017})}\BibitemShut {NoStop}%
\bibitem [{\citenamefont {Lanco}\ \emph {et~al.}(2006)\citenamefont {Lanco},
  \citenamefont {Ducci}, \citenamefont {Likforman}, \citenamefont {Marcadet},
  \citenamefont {van Houwelingen}, \citenamefont {Zbinden}, \citenamefont
  {Leo},\ and\ \citenamefont {Berger}}]{Lanco2006}%
  \BibitemOpen
  \bibfield  {author} {\bibinfo {author} {\bibfnamefont {L.}~\bibnamefont
  {Lanco}}, \bibinfo {author} {\bibfnamefont {S.}~\bibnamefont {Ducci}},
  \bibinfo {author} {\bibfnamefont {J.-P.}\ \bibnamefont {Likforman}}, \bibinfo
  {author} {\bibfnamefont {X.}~\bibnamefont {Marcadet}}, \bibinfo {author}
  {\bibfnamefont {J.~A.~W.}\ \bibnamefont {van Houwelingen}}, \bibinfo {author}
  {\bibfnamefont {H.}~\bibnamefont {Zbinden}}, \bibinfo {author} {\bibfnamefont
  {G.}~\bibnamefont {Leo}}, \ and\ \bibinfo {author} {\bibfnamefont
  {V.}~\bibnamefont {Berger}},\ }\bibfield  {title} {\enquote {\bibinfo {title}
  {Semiconductor waveguide source of counterpropagating twin photons},}\
  }\href@noop {} {\bibfield  {journal} {\bibinfo  {journal} {Phys. Rev. Lett.}\
  }\textbf {\bibinfo {volume} {97}},\ \bibinfo {pages} {173901}(\bibinfo {year} {2006})}\BibitemShut
  {NoStop}%
\bibitem [{\citenamefont {Ravaro}\ \emph {et~al.}(2005)\citenamefont {Ravaro},
  \citenamefont {Seurin}, \citenamefont {Ducci}, \citenamefont {Leo},
  \citenamefont {Berger}, \citenamefont {Rossi},\ and\ \citenamefont
  {Assanto}}]{Ravaro2005}%
  \BibitemOpen
  \bibfield  {author} {\bibinfo {author} {\bibfnamefont {M.}~\bibnamefont
  {Ravaro}}, \bibinfo {author} {\bibfnamefont {Y.}~\bibnamefont {Seurin}},
  \bibinfo {author} {\bibfnamefont {S.}~\bibnamefont {Ducci}}, \bibinfo
  {author} {\bibfnamefont {G.}~\bibnamefont {Leo}}, \bibinfo {author}
  {\bibfnamefont {V.}~\bibnamefont {Berger}}, \bibinfo {author} {\bibfnamefont
  {A.~De}\ \bibnamefont {Rossi}}, \ and\ \bibinfo {author} {\bibfnamefont
  {G.}~\bibnamefont {Assanto}},\ }\bibfield  {title} {\enquote {\bibinfo
  {title} {Nonlinear {AlGaAs} waveguide for the generation of
  counterpropagating twin photons in the telecom range},}\ }\href {\doibase
  10.1063/1.2058197} {\bibfield  {journal} {\bibinfo  {journal} {J. Appl.
  Phys.}\ }\textbf {\bibinfo {volume} {98}},\ \bibinfo {pages} {063103}
  (\bibinfo {year} {2005})}\BibitemShut {NoStop}%
\bibitem [{\citenamefont {Canalias}\ and\ \citenamefont
  {Pasiskevicius}(2007)}]{Canalias2007}%
  \BibitemOpen
  \bibfield  {author} {\bibinfo {author} {\bibfnamefont {C.}\ \bibnamefont
  {Canalias}}\ and\ \bibinfo {author} {\bibfnamefont {V.}\ \bibnamefont
  {Pasiskevicius}},\ }\bibfield  {title} {\enquote {\bibinfo {title}
  {Mirrorless optical parametric oscillator},}\ }\href {\doibase
  10.1038/nphoton.2007.137} {\bibfield  {journal} {\bibinfo  {journal} {Nat.
  Photonics}\ }\textbf {\bibinfo {volume} {1}},\ \bibinfo {pages} {459--462}
  (\bibinfo {year} {2007})}\BibitemShut {NoStop}%
\bibitem [{\citenamefont {Liu}\ \emph {et~al.}()\citenamefont {Liu},
  \citenamefont {Guo}, \citenamefont {Yang}, \citenamefont {Sun}, \citenamefont
  {Duan}, \citenamefont {Xie}, \citenamefont {Gong},\ and\ \citenamefont
  {Zhu}}]{Liu2019}%
  \BibitemOpen
  \bibfield  {author} {\bibinfo {author} {\bibfnamefont {Y.-C.}\ \bibnamefont
  {Liu}}, \bibinfo {author} {\bibfnamefont {D.-J.}\ \bibnamefont {Guo}},
  \bibinfo {author} {\bibfnamefont {R.}\ \bibnamefont {Yang}}, \bibinfo
  {author} {\bibfnamefont {C.-W.}\ \bibnamefont {Sun}}, \bibinfo {author}
  {\bibfnamefont {J.-C.}\ \bibnamefont {Duan}}, \bibinfo {author}
  {\bibfnamefont {Z.}\ \bibnamefont {Xie}}, \bibinfo {author}
  {\bibfnamefont {Y.-X.}\ \bibnamefont {Gong}}, \ and\ \bibinfo {author}
  {\bibfnamefont {S.-N.}\ \bibnamefont {Zhu}},\ }\bibfield  {title}
  {\enquote {\bibinfo {title} {Narrow-band photonic quantum entanglement with
  counterpropagating domain engineering},}\ }\href@noop {} {\bibinfo  {journal}
  {arXiv:1905.13395}\ }\BibitemShut {NoStop}%
\bibitem [{\citenamefont {Cheng}\ \emph {et~al.}(1994)\citenamefont {Cheng},
  \citenamefont {Cheng}, \citenamefont {Galperin}, \citenamefont
  {Hotsenpiller},\ and\ \citenamefont {Bierlein}}]{Cheng1994}%
  \BibitemOpen
\bibfield  {journal} {  }\bibfield  {author} {\bibinfo {author} {\bibfnamefont
  {L.}\ \bibnamefont {Cheng}}, \bibinfo {author} {\bibfnamefont {L.}\
  \bibnamefont {Cheng}}, \bibinfo {author} {\bibfnamefont {J.}~\bibnamefont
  {Galperin}}, \bibinfo {author} {\bibfnamefont {P.}\ \bibnamefont
  {Hotsenpiller}}, \ and\ \bibinfo {author} {\bibfnamefont {J.}\ \bibnamefont
  {Bierlein}},\ }\bibfield  {title} {\enquote {\bibinfo {title} {Crystal growth
  and characterization of {KTiOPO$_4$} isomorphs from the self-fluxes},}\
  }\href {\doibase 10.1016/0022-0248(94)91256-4} {\bibfield  {journal}
  {\bibinfo  {journal} {J. Cryst. Growth}\ }\textbf {\bibinfo {volume} {137}},\
  \bibinfo {pages} {107--115} (\bibinfo {year} {1994})}\BibitemShut {NoStop}%
\bibitem [{\citenamefont {Shur}\ \emph {et~al.}(2016)\citenamefont {Shur},
  \citenamefont {Pelegova}, \citenamefont {Akhmatkhanov},\ and\ \citenamefont
  {Baturin}}]{Shur2016}%
  \BibitemOpen
  \bibfield  {author} {\bibinfo {author} {\bibfnamefont {V. Y.}\ \bibnamefont
  {Shur}}, \bibinfo {author} {\bibfnamefont {E. V.}\ \bibnamefont {Pelegova}},
  \bibinfo {author} {\bibfnamefont {A. R.}\ \bibnamefont {Akhmatkhanov}}, \
  and\ \bibinfo {author} {\bibfnamefont {I. S.}\ \bibnamefont {Baturin}},\
  }\bibfield  {title} {\enquote {\bibinfo {title} {Periodically poled crystals
  of {KTP} family: a review},}\ }\href {\doibase 10.1080/00150193.2016.1157437}
  {\bibfield  {journal} {\bibinfo  {journal} {Ferr.}\ }\textbf {\bibinfo
  {volume} {496}},\ \bibinfo {pages} {49--69} (\bibinfo {year}
  {2016})}\BibitemShut {NoStop}%
\bibitem [{\citenamefont {Jin}\ \emph {et~al.}(2016)\citenamefont {Jin},
  \citenamefont {Zhao}, \citenamefont {Deng},\ and\ \citenamefont
  {Wu}}]{Jin2016PRAppl}%
  \BibitemOpen
  \bibfield  {author} {\bibinfo {author} {\bibfnamefont {R.-B.}\ \bibnamefont
  {Jin}}, \bibinfo {author} {\bibfnamefont {P.}\ \bibnamefont {Zhao}},
  \bibinfo {author} {\bibfnamefont {P.}\ \bibnamefont {Deng}}, \ and\
  \bibinfo {author} {\bibfnamefont {Q.-L.}\ \bibnamefont {Wu}},\ }\bibfield
  {title} {\enquote {\bibinfo {title} {Spectrally pure states at
  telecommunications wavelengths from periodically poled {MTiOXO$_4$ (M = K,
  Rb, Cs; X = P, As)} crystals},}\ }\href
  {https://link.aps.org/doi/10.1103/PhysRevApplied.6.064017} {\bibfield
  {journal} {\bibinfo  {journal} {Phys. Rev. Appl.}\ }\textbf {\bibinfo
  {volume} {6}},\ \bibinfo {pages} {064017} (\bibinfo {year}
  {2016})}\BibitemShut {NoStop}%
\bibitem [{\citenamefont {Laudenbach}\ \emph {et~al.}(2017)\citenamefont
  {Laudenbach}, \citenamefont {Jin}, \citenamefont {Greganti}, \citenamefont
  {Hentschel}, \citenamefont {Walther},\ and\ \citenamefont
  {H{\"u}bel}}]{Laudenbach2017}%
  \BibitemOpen
  \bibfield  {author} {\bibinfo {author} {\bibfnamefont {F.}\ \bibnamefont
  {Laudenbach}}, \bibinfo {author} {\bibfnamefont {R.-B.}\ \bibnamefont
  {Jin}}, \bibinfo {author} {\bibfnamefont {C.}\ \bibnamefont {Greganti}},
  \bibinfo {author} {\bibfnamefont {M.}\ \bibnamefont {Hentschel}},
  \bibinfo {author} {\bibfnamefont {P.}\ \bibnamefont {Walther}}, \ and\
  \bibinfo {author} {\bibfnamefont {H.}\ \bibnamefont {H{\"u}bel}},\
  }\bibfield  {title} {\enquote {\bibinfo {title} {Numerical investigation of
  photon-pair generation in periodically poled {MTiOXO$_4$ (M = K, Rb, Cs; X =
  P, As)}},}\ }\href@noop {} {\bibfield  {journal} {\bibinfo  {journal} {Phys.
  Rev. Appl.}\ }\textbf {\bibinfo {volume} {8}},\ \bibinfo {pages} {024035}
  (\bibinfo {year} {2017})}\BibitemShut {NoStop}%
\bibitem [{\citenamefont {Jin}\ \emph {et~al.}(2019)\citenamefont {Jin},
  \citenamefont {Chen}, \citenamefont {Laudenbach}, \citenamefont {Zhao},\ and\
  \citenamefont {Lu}}]{Jin2019OLT}%
  \BibitemOpen
  \bibfield  {author} {\bibinfo {author} {\bibfnamefont {R.-B.}\ \bibnamefont
  {Jin}}, \bibinfo {author} {\bibfnamefont {G.-Q.}\ \bibnamefont {Chen}},
  \bibinfo {author} {\bibfnamefont {F.}\ \bibnamefont {Laudenbach}},
  \bibinfo {author} {\bibfnamefont {S.}\ \bibnamefont {Zhao}}, \ and\
  \bibinfo {author} {\bibfnamefont {P.-X.}\ \bibnamefont {Lu}},\ }\bibfield
   {title} {\enquote {\bibinfo {title} {Thermal effects of the quantum states
  generated from the isomorphs of {PPKTP} crystal},}\ }\href@noop {} {\bibfield
   {journal} {\bibinfo  {journal} {Opt. Laser Technol.}\ }\textbf {\bibinfo
  {volume} {109}},\ \bibinfo {pages} {222--226} (\bibinfo {year}
  {2019})}\BibitemShut {NoStop}%
\bibitem [{\citenamefont {Kim}\ \emph {et~al.}(2015)\citenamefont {Kim},
  \citenamefont {Lee}, \citenamefont {Lee},\ and\ \citenamefont
  {Moon}}]{Kim2015}%
  \BibitemOpen
  \bibfield  {author} {\bibinfo {author} {\bibfnamefont {H.}\ \bibnamefont
  {Kim}}, \bibinfo {author} {\bibfnamefont {H. J.}\ \bibnamefont {Lee}},
  \bibinfo {author} {\bibfnamefont {S. M.}\ \bibnamefont {Lee}}, \ and\
  \bibinfo {author} {\bibfnamefont {H. S.}\ \bibnamefont {Moon}},\ }\bibfield
   {title} {\enquote {\bibinfo {title} {Highly efficient source for
  frequency-entangled photon pairs generated in a 3$^{rd}$-order periodically
  poled {MgO}-doped stoichiometric {LiTaO$_3$} crystal},}\ }\href {\doibase
  10.1364/ol.40.003061} {\bibfield  {journal} {\bibinfo  {journal} {Opt.
  Lett.}\ }\textbf {\bibinfo {volume} {40}},\ \bibinfo {pages} {3061--3064} (\bibinfo
  {year} {2015})}\BibitemShut {NoStop}%
\bibitem [{\citenamefont {Mosley}\ \emph
  {et~al.}(2008{\natexlab{b}})\citenamefont {Mosley}, \citenamefont {Lundeen},
  \citenamefont {Smith},\ and\ \citenamefont {Walmsley}}]{Mosley2008NJP}%
  \BibitemOpen
  \bibfield  {author} {\bibinfo {author} {\bibfnamefont {P. J.}\ \bibnamefont
  {Mosley}}, \bibinfo {author} {\bibfnamefont {J. S.}\ \bibnamefont {Lundeen}},
  \bibinfo {author} {\bibfnamefont {B. J.}\ \bibnamefont {Smith}}, \ and\
  \bibinfo {author} {\bibfnamefont {I. A.}\ \bibnamefont {Walmsley}},\ }\bibfield
   {title} {\enquote {\bibinfo {title} {Conditional preparation of single
  photons using parametric downconversion: a recipe for purity},}\ }\href@noop
  {} {\bibfield  {journal} {\bibinfo  {journal} {New J. Phys.}\ }\textbf
  {\bibinfo {volume} {10}},\ \bibinfo {pages} {093011} (\bibinfo {year}
  {2008}{\natexlab{b}})}\BibitemShut {NoStop}%
\bibitem [{\citenamefont {Jin}\ \emph {et~al.}(2013{\natexlab{a}})\citenamefont
  {Jin}, \citenamefont {Shimizu}, \citenamefont {Wakui}, \citenamefont
  {Benichi},\ and\ \citenamefont {Sasaki}}]{Jin2013}%
  \BibitemOpen
  \bibfield  {author} {\bibinfo {author} {\bibfnamefont {R.-B.}\ \bibnamefont
  {Jin}}, \bibinfo {author} {\bibfnamefont {R.}\ \bibnamefont {Shimizu}},
  \bibinfo {author} {\bibfnamefont {K.}\ \bibnamefont {Wakui}}, \bibinfo
  {author} {\bibfnamefont {H.}\ \bibnamefont {Benichi}}, \ and\ \bibinfo
  {author} {\bibfnamefont {M.}\ \bibnamefont {Sasaki}},\ }\bibfield
  {title} {\enquote {\bibinfo {title} {Widely tunable single photon source with
  high purity at telecom wavelength},}\ }\href@noop {} {\bibfield  {journal}
  {\bibinfo  {journal} {Opt. Express}\ }\textbf {\bibinfo {volume} {21}},\
  \bibinfo {pages} {10659--10666} (\bibinfo {year} {2013}{\natexlab{a}})}\BibitemShut
  {NoStop}%
\bibitem [{\citenamefont {Kato}\ and\ \citenamefont
  {Takaoka}(2002)}]{Kato2002}%
  \BibitemOpen
  \bibfield  {author} {\bibinfo {author} {\bibfnamefont {K.}\ \bibnamefont
  {Kato}}\ and\ \bibinfo {author} {\bibfnamefont {E.}\ \bibnamefont
  {Takaoka}},\ }\bibfield  {title} {\enquote {\bibinfo {title} {Sellmeier and
  thermo-optic dispersion formulas for {KTP}},}\ }\href {\doibase
  10.1364/ao.41.005040} {\bibfield  {journal} {\bibinfo  {journal} {Appl.
  Opt.}\ }\textbf {\bibinfo {volume} {41}},\ \bibinfo {pages} {5040--5044} (\bibinfo
  {year} {2002})}\BibitemShut {NoStop}%
\bibitem [{\citenamefont {Mikami}\ \emph {et~al.}(2009)\citenamefont {Mikami},
  \citenamefont {Okamoto},\ and\ \citenamefont {Kato}}]{Mikami2009}%
  \BibitemOpen
  \bibfield  {author} {\bibinfo {author} {\bibfnamefont {T.}\ \bibnamefont
  {Mikami}}, \bibinfo {author} {\bibfnamefont {T.}\ \bibnamefont
  {Okamoto}}, \ and\ \bibinfo {author} {\bibfnamefont {K.}\ \bibnamefont
  {Kato}},\ }\bibfield  {title} {\enquote {\bibinfo {title} {Sellmeier and
  thermo-optic dispersion formulas for {RbTiOPO$_4$}},}\ }\href {\doibase
  10.1016/j.optmat.2009.03.012} {\bibfield  {journal} {\bibinfo  {journal}
  {Opt. Mater.}\ }\textbf {\bibinfo {volume} {31}},\ \bibinfo {pages}
  {1628--1630} (\bibinfo {year} {2009})}\BibitemShut {NoStop}%
\bibitem [{\citenamefont {F{\"e}ve}\ \emph {et~al.}(2000)\citenamefont
  {F{\"e}ve}, \citenamefont {Boulanger}, \citenamefont {Pacaud}, \citenamefont
  {Rousseau}, \citenamefont {M{\"e}naert}, \citenamefont {Marnier},
  \citenamefont {Villeval}, \citenamefont {Bonnin}, \citenamefont {Loiacono},\
  and\ \citenamefont {Loiacono}}]{Feve2000}%
  \BibitemOpen
  \bibfield  {author} {\bibinfo {author} {\bibfnamefont {J. -P.}\
  \bibnamefont {F{\"e}ve}}, \bibinfo {author} {\bibfnamefont {B.}\
  \bibnamefont {Boulanger}}, \bibinfo {author} {\bibfnamefont {O.}\
  \bibnamefont {Pacaud}}, \bibinfo {author} {\bibfnamefont {I.}\
  \bibnamefont {Rousseau}}, \bibinfo {author} {\bibfnamefont {B.}\
  \bibnamefont {M{\"e}naert}}, \bibinfo {author} {\bibfnamefont {G.}\
  \bibnamefont {Marnier}}, \bibinfo {author} {\bibfnamefont {P.}\
  \bibnamefont {Villeval}}, \bibinfo {author} {\bibfnamefont {C.}\
  \bibnamefont {Bonnin}}, \bibinfo {author} {\bibfnamefont {G. M.}\
  \bibnamefont {Loiacono}}, \ and\ \bibinfo {author} {\bibfnamefont {D. N.}\
  \bibnamefont {Loiacono}},\ }\bibfield  {title} {\enquote {\bibinfo {title}
  {Phase-matching measurements and sellmeier equations over the complete
  transparency range of {KTiOAsO$_4$}, {RbTiOAsO$_4$}, and {CsTiOAsO$_4$}},}\
  }\href {\doibase 10.1364/josab.17.000775} {\bibfield  {journal} {\bibinfo
  {journal} {J. Opt. Soc. Am. B}\ }\textbf {\bibinfo {volume} {17}},\ \bibinfo
  {pages} {775--780} (\bibinfo {year} {2000})}\BibitemShut {NoStop}%
\bibitem [{\citenamefont {Kato}\ \emph {et~al.}(2003)\citenamefont {Kato},
  \citenamefont {Takaoka},\ and\ \citenamefont {Umemura}}]{Kato2003}%
  \BibitemOpen
  \bibfield  {author} {\bibinfo {author} {\bibfnamefont {K.}\ \bibnamefont
  {Kato}}, \bibinfo {author} {\bibfnamefont {E.}\ \bibnamefont {Takaoka}}, \
  and\ \bibinfo {author} {\bibfnamefont {N.}\ \bibnamefont {Umemura}},\
  }\bibfield  {title} {\enquote {\bibinfo {title} {Thermo-optic dispersion
  formula for {RbTiOAsO$_4$}},}\ }\href {\doibase 10.1143/jjap.42.6420}
  {\bibfield  {journal} {\bibinfo  {journal} {Jpn. Appl. Phys.}\ }\textbf
  {\bibinfo {volume} {42}},\ \bibinfo {pages} {6420--6423} (\bibinfo {year}
  {2003})}\BibitemShut {NoStop}%
\bibitem [{\citenamefont {Mikami}\ \emph {et~al.}(2011)\citenamefont {Mikami},
  \citenamefont {Okamoto},\ and\ \citenamefont {Kato}}]{Mikami2011}%
  \BibitemOpen
  \bibfield  {author} {\bibinfo {author} {\bibfnamefont {T.}\ \bibnamefont
  {Mikami}}, \bibinfo {author} {\bibfnamefont {T.}\ \bibnamefont
  {Okamoto}}, \ and\ \bibinfo {author} {\bibfnamefont {K.}\ \bibnamefont
  {Kato}},\ }\bibfield  {title} {\enquote {\bibinfo {title} {Sellmeier and
  thermo-optic dispersion formulas for {CsTiOAsO$_4$}},}\ }\href {\doibase
  10.1063/1.3525800} {\bibfield  {journal} {\bibinfo  {journal} {J. Appl.
  Phys.}\ }\textbf {\bibinfo {volume} {109}},\ \bibinfo {pages} {023108}
  (\bibinfo {year} {2011})}\BibitemShut {NoStop}%
\bibitem [{\citenamefont {Smith}()}]{SNLO70}%
  \BibitemOpen
  \bibfield  {author} {\bibinfo {author} {\bibfnamefont {A.}~\bibnamefont
  {Smith}},\ }\href@noop {} {\enquote {\bibinfo {title} {Snlo},}\ }\bibinfo
  {howpublished} {http://www.as-photonics.com/snlo}\BibitemShut {NoStop}%
\bibitem [{\citenamefont {Hong}\ \emph {et~al.}(1987)\citenamefont {Hong},
  \citenamefont {Ou},\ and\ \citenamefont {Mandel}}]{Hong1987}%
  \BibitemOpen
  \bibfield  {author} {\bibinfo {author} {\bibfnamefont {C. K.}\ \bibnamefont
  {Hong}}, \bibinfo {author} {\bibfnamefont {Z. Y.}\ \bibnamefont {Ou}}, \ and\
  \bibinfo {author} {\bibfnamefont {L.}~\bibnamefont {Mandel}},\ }\bibfield
  {title} {\enquote {\bibinfo {title} {Measurement of subpicosecond time
  intervals between two photons by interference},}\ }\href {\doibase
  10.1103/physrevlett.59.2044} {\bibfield  {journal} {\bibinfo  {journal}
  {Phys. Rev. Lett.}\ }\textbf {\bibinfo {volume} {59}},\ \bibinfo {pages}
  {2044--2046} (\bibinfo {year} {1987})}\BibitemShut {NoStop}%
\bibitem [{\citenamefont {Jin}\ \emph {et~al.}(2013{\natexlab{b}})\citenamefont
  {Jin}, \citenamefont {Wakui}, \citenamefont {Shimizu}, \citenamefont
  {Benichi}, \citenamefont {Miki}, \citenamefont {Yamashita}, \citenamefont
  {Terai}, \citenamefont {Wang}, \citenamefont {Fujiwara},\ and\ \citenamefont
  {Sasaki}}]{Jin2013PRA}%
  \BibitemOpen
  \bibfield  {author} {\bibinfo {author} {\bibfnamefont {R.-B.}\ \bibnamefont
  {Jin}}, \bibinfo {author} {\bibfnamefont {K.}\ \bibnamefont {Wakui}},
  \bibinfo {author} {\bibfnamefont {R.}\ \bibnamefont {Shimizu}}, \bibinfo
  {author} {\bibfnamefont {H.}\ \bibnamefont {Benichi}}, \bibinfo {author}
  {\bibfnamefont {S.}\ \bibnamefont {Miki}}, \bibinfo {author}
  {\bibfnamefont {T.}\ \bibnamefont {Yamashita}}, \bibinfo {author}
  {\bibfnamefont {H.}\ \bibnamefont {Terai}}, \bibinfo {author}
  {\bibfnamefont {Z.}\ \bibnamefont {Wang}}, \bibinfo {author} {\bibfnamefont
  {M.}\ \bibnamefont {Fujiwara}}, \ and\ \bibinfo {author} {\bibfnamefont
  {M.}\ \bibnamefont {Sasaki}},\ }\bibfield  {title} {\enquote {\bibinfo
  {title} {Nonclassical interference between independent intrinsically pure
  single photons at telecommunication wavelength},}\ }\href@noop {} {\bibfield
  {journal} {\bibinfo  {journal} {Phys. Rev. A}\ }\textbf {\bibinfo {volume}
  {87}},\ \bibinfo {pages} {063801} (\bibinfo {year} {2013}{\natexlab{b}})}\BibitemShut {NoStop}%
\bibitem [{\citenamefont {Grice}\ and\ \citenamefont
  {Walmsley}(1997)}]{Grice1997}%
  \BibitemOpen
  \bibfield  {author} {\bibinfo {author} {\bibfnamefont {W.~P.}\ \bibnamefont
  {Grice}}\ and\ \bibinfo {author} {\bibfnamefont {I.~A.}\ \bibnamefont
  {Walmsley}},\ }\bibfield  {title} {\enquote {\bibinfo {title} {Spectral
  information and distinguishability in type-{II} down-conversion with a
  broadband pump},}\ }\href {\doibase 10.1103/physreva.56.1627} {\bibfield
  {journal} {\bibinfo  {journal} {Phys. Rev. A}\ }\textbf {\bibinfo {volume}
  {56}},\ \bibinfo {pages} {1627--1634} (\bibinfo {year} {1997})}\BibitemShut
  {NoStop}%
\bibitem [{\citenamefont {Ou}(2007)}]{Ou2007}%
  \BibitemOpen
  \bibfield  {author} {\bibinfo {author} {\bibfnamefont {Z.-Y. J.}\
  \bibnamefont {Ou}},\ }\href {\doibase 10.1007/978-0-387-25554-5} {\emph
  {\bibinfo {title} {Multi-Photon Quantum Interference}}}\ (\bibinfo
  {publisher} {Springer},\ \bibinfo {year} {2007})\BibitemShut {NoStop}%
\bibitem [{\citenamefont {Jin}\ \emph {et~al.}(2020)\citenamefont {Jin},
  \citenamefont {Cai}, \citenamefont {Ding}, \citenamefont {Mei}, \citenamefont
  {Deng}, \citenamefont {Shimizu},\ and\ \citenamefont {Zhou}}]{Jin2020QUE}%
  \BibitemOpen
  \bibfield  {author} {\bibinfo {author} {\bibfnamefont {R.-B.}\ \bibnamefont
  {Jin}}, \bibinfo {author} {\bibfnamefont {W.-H.}\ \bibnamefont {Cai}},
  \bibinfo {author} {\bibfnamefont {C.}\ \bibnamefont {Ding}}, \bibinfo
  {author} {\bibfnamefont {F.}\ \bibnamefont {Mei}}, \bibinfo {author}
  {\bibfnamefont {G.-W.}\ \bibnamefont {Deng}}, \bibinfo {author}
  {\bibfnamefont {R.}\ \bibnamefont {Shimizu}}, \ and\ \bibinfo {author}
  {\bibfnamefont {Q.}\ \bibnamefont {Zhou}},\ }\bibfield  {title} {\enquote
  {\bibinfo {title} {Spectrally uncorrelated biphotons generated from 'the
  family of {BBO} crystal'},}\ }\href@noop {} {\bibfield  {journal} {\bibinfo
  {journal} {Quantum Eng.}\ }\textbf {\bibinfo {volume} {2}},\ \bibinfo {pages} {e38} (\bibinfo {year}
  {2020})}\BibitemShut {NoStop}%
\bibitem [{\citenamefont {Cui}\ \emph {et~al.}(2019)\citenamefont {Cui},
  \citenamefont {Arian}, \citenamefont {Guha}, \citenamefont {Peyghambarian},
  \citenamefont {Zhuang},\ and\ \citenamefont {Zhang}}]{Cui2019PRAppl}%
  \BibitemOpen
  \bibfield  {author} {\bibinfo {author} {\bibfnamefont {C.}\ \bibnamefont
  {Cui}}, \bibinfo {author} {\bibfnamefont {R.}\ \bibnamefont {Arian}},
  \bibinfo {author} {\bibfnamefont {S.}\ \bibnamefont {Guha}}, \bibinfo
  {author} {\bibfnamefont {N.}~\bibnamefont {Peyghambarian}}, \bibinfo {author}
  {\bibfnamefont {Q.}\ \bibnamefont {Zhuang}}, \ and\ \bibinfo {author}
  {\bibfnamefont {Z.}\ \bibnamefont {Zhang}},\ }\bibfield  {title}
  {\enquote {\bibinfo {title} {Wave-function engineering for spectrally
  uncorrelated biphotons in the telecommunication band based on a
  machine-learning framework},}\ }\href@noop {} {\bibfield  {journal} {\bibinfo
   {journal} {Phys. Rev. Appl.}\ }\textbf {\bibinfo {volume} {12}},\ \bibinfo
  {pages} {034059} (\bibinfo {year} {2019})}\BibitemShut {NoStop}%
\bibitem [{\citenamefont {Chen}\ \emph {et~al.}(2019)\citenamefont {Chen},
  \citenamefont {Heyes}, \citenamefont {Hong}, \citenamefont {Niu},
  \citenamefont {Lita}, \citenamefont {Gerrits}, \citenamefont {Nam},
  \citenamefont {Shapiro},\ and\ \citenamefont {Wong}}]{Chen2019OE}%
  \BibitemOpen
  \bibfield  {author} {\bibinfo {author} {\bibfnamefont {C.}\
  \bibnamefont {Chen}}, \bibinfo {author} {\bibfnamefont {J. E.}\
  \bibnamefont {Heyes}}, \bibinfo {author} {\bibfnamefont {K.-H.}\
  \bibnamefont {Hong}}, \bibinfo {author} {\bibfnamefont {M. Y.}\
  \bibnamefont {Niu}}, \bibinfo {author} {\bibfnamefont {A. E.}\
  \bibnamefont {Lita}}, \bibinfo {author} {\bibfnamefont {T.}\ \bibnamefont
  {Gerrits}}, \bibinfo {author} {\bibfnamefont {S. W.}\ \bibnamefont {Nam}},
  \bibinfo {author} {\bibfnamefont {J. H.}\ \bibnamefont {Shapiro}}, \
  and\ \bibinfo {author} {\bibfnamefont {F. N. C.}\ \bibnamefont {Wong}},\
  }\bibfield  {title} {\enquote {\bibinfo {title} {Indistinguishable
  single-mode photons from spectrally engineered biphotons},}\ }\href {\doibase
  10.1364/oe.27.011626} {\bibfield  {journal} {\bibinfo  {journal} {Opt.
  Express}\ }\textbf {\bibinfo {volume} {27}},\ \bibinfo {pages} {11626--11634}
  (\bibinfo {year} {2019})}\BibitemShut {NoStop}%
\bibitem [{\citenamefont {Graffitti}\ \emph {et~al.}(2018)\citenamefont
  {Graffitti}, \citenamefont {Barrow}, \citenamefont {Proietti}, \citenamefont
  {Kundys},\ and\ \citenamefont {Fedrizzi}}]{Graffitti2018Optica}%
  \BibitemOpen
  \bibfield  {author} {\bibinfo {author} {\bibfnamefont {F.}\
  \bibnamefont {Graffitti}}, \bibinfo {author} {\bibfnamefont {P.}\
  \bibnamefont {Barrow}}, \bibinfo {author} {\bibfnamefont {M.}\
  \bibnamefont {Proietti}}, \bibinfo {author} {\bibfnamefont {D.}\
  \bibnamefont {Kundys}}, \ and\ \bibinfo {author} {\bibfnamefont {A.}\
  \bibnamefont {Fedrizzi}},\ }\bibfield  {title} {\enquote {\bibinfo {title}
  {Independent high-purity photons created in domain-engineered crystals},}\
  }\href {\doibase 10.1364/optica.5.000514} {\bibfield  {journal} {\bibinfo
  {journal} {Optica}\ }\textbf {\bibinfo {volume} {5}},\ \bibinfo {pages} {514--517}
  (\bibinfo {year} {2018})}\BibitemShut {NoStop}%
\end{thebibliography}
\end{document}